\documentclass[review]{elsarticle}

\usepackage{lineno,hyperref}
\usepackage{subcaption}
\usepackage{listings}
\usepackage{algpseudocode}
\usepackage[ruled,vlined]{algorithm2e}

\SetCommentSty{mycommfont}
\usepackage{lineno,hyperref,xcolor}
\SetKwInput{KwInput}{Input}                
\SetKwInput{KwOutput}{Output} 

\usepackage{enumerate}
\usepackage{bibentry}
\usepackage{graphicx,xcolor,multicol,eso-pic,mwe,hyphenat}
\usepackage{tcolorbox}[ 
	colback=White
	]
\usepackage{epstopdf}
\usepackage{caption,subcaption,rotating,pdflscape,float}
\usepackage{siunitx} 
\usepackage[version=4]{mhchem}
\mhchemoptions{version=4}

\usepackage{hyperref} 

\usepackage{pgfplots}
\usepackage{tikz}
\usetikzlibrary{shapes.multipart}
\usetikzlibrary{shapes.geometric,arrows} 
\pgfplotsset{compat=1.9}

\pgfplotsset{
    myplotstyle/.style={
    legend style={draw=none, font=\small},
    legend cell align=left,
    legend pos=north east,
    ylabel style={align=center},
    xlabel style={align=center},
    x tick label style={},
    y tick label style={},
    xtick pos=left, 
    ytick pos=left,
    scaled ticks=false,
    every axis plot/.append style={thick},
    },
}
\usepackage{amsmath, multirow,amssymb,amsthm,textcomp}
\usepackage{tikz,amsmath}
\usetikzlibrary{matrix,calc}
\usetikzlibrary{shapes.multipart}
\usetikzlibrary{shapes.geometric,arrows}
\usepackage{multirow}
\usepackage{adjustbox}
\usepackage{array}
\usepackage{setspace}
\usepackage[section]{placeins}

\journal{Medical Image Analysis}

\begin{document}

\begin{frontmatter}

\title{Detection of Under-represented Samples Using Dynamic Batch Training for Brain Tumor Segmentation from MR Images}


\author[1]{Subin Sahayam}
\corref{cor1}
\ead{subinsahayamm@snuchennai.edu.in, coe18d001@iiitdm.ac.in}

\address[1]{Department of Computer Science and Engineering, Shiv Nadar University Chennai, Kalavakkam - 603110, Tamil Nadu, India}

\address[2]{Department of Computer Science and Engineering, Indian Institute of Information Technology Design and Manufacturing Kancheepuram, Chennai - 600127, Tamil Nadu, India}

\author[2]{John Michael Sujay Zakkam}
\ead{ced18i059@iiitdm.ac.in}

\author[2]{Yoga Sri Varshan V}
\ead{ced18i058@iiitdm.ac.in}

\author[2]{Umarani Jayaraman}
\ead{umarani@iiitdm.ac.in}

\cortext[cor1]{Corresponding author}

\begin{abstract}
Brain tumors in magnetic resonance imaging (MR) are difficult, time-consuming, and prone to human error. These challenges can be resolved by developing automatic brain tumor segmentation methods from MR images. Various deep-learning models based on the U-Net have been proposed for the task. These deep-learning models are trained on a dataset of tumor images and then used for segmenting the masks. Mini-batch training is a widely used method in deep learning for training. However, one of the significant challenges associated with this approach is that if the training dataset has under-represented samples or samples with complex latent representations, the model may not generalize well to these samples. The issue leads to skewed learning of the data, where the model learns to fit towards the majority representations while underestimating the under-represented samples. The proposed dynamic batch training method addresses the challenges posed by under-represented data points, data points with complex latent representation, and imbalances within the class, where some samples may be harder to learn than others. Poor performance of such samples can be identified only after the completion of the training, leading to the wastage of computational resources. Also, training easy samples after each epoch is an inefficient utilization of computation resources. To overcome these challenges, the proposed method identifies hard samples and trains such samples for more iterations compared to easier samples on the BraTS2020 dataset. Additionally, the samples trained multiple times are identified and it provides a way to identify hard samples in the BraTS2020 dataset. The comparison of the proposed training approach with U-Net and other models in the literature highlights the capabilities of the proposed training approach.
\end{abstract}

\begin{keyword}
Brain Tumor Segmentation \sep BraTS2020 \sep Representation Learning \sep Outlier Analysis
\end{keyword}

\end{frontmatter}
\section{Introduction}
\label{chap: intro}
 Brain tumors are abnormal growths of cells within the brain or surrounding tissues \cite{hanahan2022hallmarks, ameisen2002origin}. Diagnosis of brain tumors typically involves a combination of medical history evaluation, physical examination, and diagnostic imaging tests. Some imaging techniques used for the diagnosis of brain tumors are magnetic resonance imaging (MRI), computed tomography (CT), positron emission tomography (PET), and single-photon emission computed tomography (SPECT) \cite{kasban2015comparative}. Among these techniques, MRI is the most commonly used imaging modality for diagnosing and monitoring brain tumors.

The detection and analysis of brain tumors play a crucial role in diagnosis and treatment planning. Clinicians and radiologists generally utilize MR images to identify and understand various parameters of the tumor. Segmentation of the tumor gives information such as tumor location, structure, volume, density, and volume of neighboring cells affected by the tumor. Such information can help doctors decide on further treatment procedures \cite{pereira2016brain}.

Manual segmentation of brain tumors is a time-consuming and labor-intensive task. The average time for manual segmentation can take 3 to 5 hours \cite{kaus2001automated}. Extensive expertise is required for accurate manual segmentation. Inter-observer variability is a challenge due to differences in the interpretation of MR images and subjective judgments of tumor edges among experts \cite{akkus2017deep}. Intra-observer variability occurs when the same expert segments the same tumor multiple times and produces slightly different results  \cite{akkus2017deep}. Lack of agreement on defining tumor edges is a challenge. The edge regions between two regions can be smooth, making it difficult to define the edges accurately. The dice score for manual segmentation generally ranges between 74-85\% \cite{menze2014multimodal}. Hence, there is a need for the automation of brain tumor segmentation.

In recent years, deep learning models have been the go-to approach for segmenting brain tumors. U-Net \cite{Unet_original} and its variants have achieved good results in the literature for semantic segmentation. Automating the task of brain tumor segmentation comes with its challenges. Some main computational challenges are low contrast, intensity inhomogeneity due to bias field distortion, varying shape, size, tumor location, artifacts present in images, cost of processing 3D images, integration of multiple MR scans, head alignment issues, differing scanning protocols, and poor edge prediction \cite{wadhwa2019review}. Among these challenges, tumor edge prediction is critical as it helps plan surgeries.  Tumor edges are often smooth and blend into the neighboring regions. Low contrast, bias-field distortion, magnetic field strength, and intensity inhomogeneity can lead to smooth edge problems \cite{cong2016modified}. Another critical challenge is in identifying the small tumor regions. They are obscured by the majority of background and normal tissues making them hard to detect. Small tumor regions if missed could lead to tumor relapse.

Deep learning models used to segment tumor images are often trained over a dataset with $n$ samples, the traditional approach uses mini-batch gradient descent. It involves dividing the dataset into $k$ sub-parts called mini-batches, where $k$ is a positive integer such that $k$ $<$ $n$ and $k$ $>$ $1$. The model is then trained by passing each mini-batch through the network, calculating the loss, propagating the gradients, and updating the weights. When the process is repeated for all the $k$ mini-batches, it is called an epoch. The $k$ batches are generally shuffled after each epoch. The model's training occurs until the model converges or until a fixed number of epochs $e$ is reached. During each epoch, all the samples in the dataset are passed through the network exactly once, with the weights updated after each mini-batch. The approach has faster training times and better generalization performance. Additionally, mini-batch gradient descent allows for more efficient use of computational and memory resources, as it is typically much faster to perform calculations on smaller batches of data than the entire dataset at once. Smaller batches also require lesser memory storage \cite{khirirat2017mini}.

However, one of the significant challenges associated with this approach is that if the training dataset has under-represented samples or samples with complex latent representations, then the model may not generalize well to these samples. The issue leads to skewed learning, where the model learns to fit towards the majority representation while underestimating the under-represented samples. Several techniques have been proposed in the literature. For example, when the data imbalance is between different classes, methods such as data augmentation of the minority class, collection of more data about the under-represented data, and specialized loss functions, such as focal loss, can be employed to handle the class imbalance \cite{nalepa2019data}. However, manual intervention to track and identify under-represented data may not be feasible for larger datasets. Also, data augmentation may not always be a viable solution, particularly in critical data such as medical images, where the creation of augmented images can result in combinations that may not be possible, leading to models that do not reflect the ground reality \cite{garcea2022data}. The issue can become even more challenging when the imbalance is within the same class, making it harder to identify hard samples. Data augmentation increases the total number of images for training, thereby leading to additional utilization of computational resources. Focal loss can handle the issue to a great extent but doesn't completely solve the problem \cite{focal_tversky_loss}.

The proposed dynamic batch training method addresses the challenges posed by under-represented data points, false positives, data points with complex latent representation, imbalances within the class, and smaller tumor regions. These challenges make some samples harder to learn than others. The authors refer to these kinds of samples in a dataset as hard samples.  It may be difficult to identify such samples even after analyzing the results obtained after training a model. Also, training easy samples after each epoch is an inefficient utilization of computation resources. To overcome these challenges, the proposed method aims to identify hard samples and trains such samples for more iterations compared to easier samples. The method aims to improve the overall performance of the deep learning model in the dice and Hausdorff95 metric.

The contributions of the proposed work are as follows,
\begin{enumerate}
	\item Developed a dynamic batch training algorithm to identify hard samples.
    \item Study the performance of dynamic batch training with other deep learning models and methods in the literature.
    \item Utilized dynamic batch training algorithm to identify potential outliers for brain tumor segmentation.
	\item Proposed modified focal losses to deal with false positives.
    \item Performed ablation study on the focal modified losses to study the impact of the proposed losses in training.
\end{enumerate}

The rest of the paper is presented as follows. Section \ref{chap: related_work} discusses various related work. Section \ref{chap: proposed_methodology} introduces the proposed methodology and gives the design justifications for the enhancements. Additionally, it discusses a method to detect outliers based on the number of times a sample has been trained using the dynamic batch training method. The dataset, performance metrics, loss functions, comparison of results with other methods, and results on the training and validation have been discussed in Section \ref{chap: results}. Finally, Section \ref{chap: conclusion} discusses the conclusions of the proposed work and gives future directions.

\section{Related Work}
\label{chap: related_work}

\subsection{Model-oriented approach}
 Model-oriented approaches focus on designing efficient algorithms to solve problems by considering the input parameters and the specific operation to achieve the desired output. Specifically in deep learning, model-oriented approaches focus on designing and refining the architecture of the neural network models to improve their performance on specific tasks. These approaches often involve exploring and improving various model architectures, such as convolutional neural networks (CNNs), recurrent neural networks (RNNs), and attention-based models, and optimizing hyperparameters such as learning rates, weight decay, and batch size to achieve better performance metric results. Model-oriented approaches also involve regularization techniques such as dropout and early stopping, which prevent overfitting and improve the generalization ability of the models. Loss functions are a component of the model-oriented approach used to define the objective function the neural network tries to optimize during training \cite{xu2018model,shlezinger2023model}. Model-oriented approaches are crucial in advancing state-of-the-art performance in various applications such as computer vision, natural language processing, and speech recognition. Such solutions to solving brain tumor segmentation are abundant in the literature. Recent advances have particularly benefited from deep learning techniques, which offer significant improvements over traditional and machine learning methods, especially in handling complex brain tumor segmentation tasks involving multimodal MRI data \cite{menze2014multimodal, bakas2018identifying}.

\subsection{Data-oriented Methods}

Data-oriented approaches in deep learning use preprocessing and manipulation of data to improve the quality and quantity of data used in the training of models to improve the performance of a model. The success of deep learning models often depends on the availability of large and high-quality datasets that represent the problem at hand \cite{wuest2016machine}. Data-oriented approaches aim to address the challenges of limited or low-quality data by employing techniques such as data augmentation, normalization, data cleaning, feature selection, outlier detection, and denoising to enhance the quality and quantity of training data. Data-oriented approaches have shown great promise in improving the performance of deep learning models and have been applied to a wide range of applications, such as computer vision, natural language processing, and speech recognition. Purely data-oriented approaches are model-agnostic \cite{shlezinger2023model}, meaning such approaches can be applied irrespective of the trained model. Innovations in data augmentation, such as advanced geometric transformations and synthetic data generation, play a critical role in training deep learning models for medical imaging, especially in enhancing the variability and quality of training datasets for brain tumor segmentation \cite{shorten2019survey, chlap2021review}.

Data augmentation is a powerful technique used in deep learning to increase the size and diversity of the training data by creating new samples from the existing data. This technique helps prevent over-fitting and improves the model's performance. Data augmentation methods include flipping, rotating, zooming, and cropping images \cite{shorten2019survey}. Flipping involves flipping an image horizontally or vertically. Rotation involves rotating an image to a certain degree. Zooming involves zooming in or out on an image, while cropping involves extracting a smaller part of an image. Other data augmentation methods include adding noise, adjusting an image's brightness, contrast, and saturation levels, and applying color jitter. These techniques can be applied to images and other data types, such as text and audio. Data augmentation can be customized to specific problems and data types, and it is an essential technique in deep learning for improving the performance of models \cite{chlap2021review}.

Normalization is an essential step in data preprocessing that aims to rescale the data to a standard range to ensure that the features have the same scale and variance. Standard normalization techniques include Min-Max scaling, Z-score normalization, and L2 normalization \cite{wu2018l1}. Data cleaning involves identifying and removing noisy, incomplete, or irrelevant data from the dataset to ensure the data quality used for training the model \cite{zhou2020knowledge}. Feature selection is another data preprocessing technique that aims to identify the most relevant features in the dataset that can be used for model training. It can help improve the model's performance by reducing the number of features, thus minimizing the risk of overfitting \cite{khalid2014survey}. Outlier detection identifies and removes outliers from the dataset that can negatively impact the model's performance. Techniques such as clustering, distance-based methods, and machine learning-based approaches can be used for outlier detection \cite{pang2021deep}. Finally, denoising removes noise from the dataset to improve the data quality used for training the model. Various denoising techniques can be employed, such as wavelet denoising, median filtering, and Kalman filtering \cite{welch1995introduction,tian2019deep}.
 
Most works in the literature so far have focused on changes to the model architecture or hyperparameter tuning. Also, when it comes to brain tumors, the edge prediction of the tumor is as essential as the tumor regions themselves. Edges of the tumor regions are difficult to delineate as the transition from healthy tissue to brain tissue regions is usually smooth \cite{bakas2018identifying}. In recent years, there have been efforts to bridge the gap by designing models that are aware of the edges \cite{liu2022shape} and models in which the edge information is infused during the learning process \cite{zhu2023brain, jiang2021novel}. However, the proposed models in the literature do not apply to other models without significant changes to the base model.

The recent work, \cite{sahayam2023can}, introduces the concept of dynamic batch training, focusing on hard samples—those with higher losses during the training process. This method selectively trains on a subset of these challenging samples to enhance the model's robustness and generalization capabilities. While effective in handling complex data distributions, this approach can lead to overfitting on the hard samples and may neglect the broader dataset nuances. In our proposed work, we aim to address these hindrances by balancing the focus on hard samples with regular training intervals, thereby mitigating the risk of overfitting and improving the model’s performance across diverse scenarios.

\section{Proposed Model}
\label{chap: proposed_methodology}

Building upon the foundational concepts introduced in \cite{sahayam2023can}, the current research extends these principles specifically to the field of deep learning for medical imaging. By applying dynamic batch training to the segmentation of brain tumors, the aim is to enhance the learning efficiency and generalization capabilities of deep neural networks. This extension addresses the specific challenges of medical imaging datasets, which often include highly imbalanced and heterogeneous data, thereby improving the efficacy of tumor detection and segmentation.

\begin{figure}[ht]
	\includegraphics[width=\linewidth]{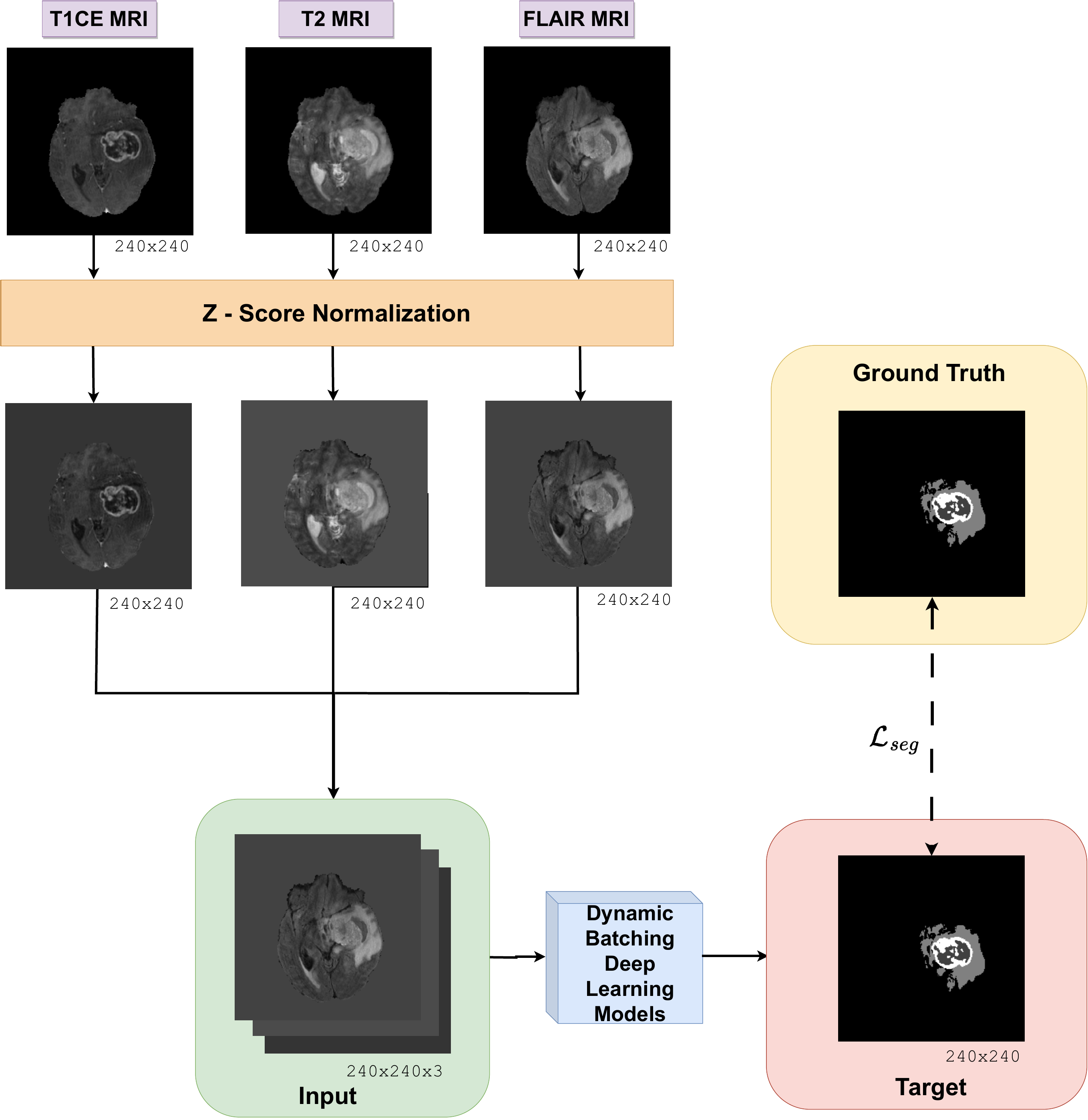}
	\caption{Proposed workflow of dynamic batch training for learning hard samples}
	\label{fig: workflow3}
\end{figure}

A novel approach to address the issue of hard samples has been proposed. Hard samples are defined as data points that have a high loss value. The idea is that models will most likely fail to predict under-represented and difficult-to-distinguish samples than numerous simple patterns. The proposed workflow is shown in Figure \ref{fig: workflow3}, it involves two significant steps: data preprocessing and dynamic batch training. In the pre-processing step, z-score normalization, one-hot vector preparation, stacking of input (T2, FLAIR, T1CE) MR images, and stacking of target one-hot segments has been done. Then, the dynamic batch training step, a technique to identify the hard samples and train them for more iterations compared to easier samples has been proposed. The goal of the approach is to make sure that hard samples are focused with attention during training and help to improve the overall performance of the model. In the following sections, a detailed description of the data pre-processing and dynamic batch training steps has been discussed.
\subsection{Data Preprocessing}
\begin{figure}[htb]
\captionsetup[subfigure]{justification=centering}
	\centering
	\subfloat[2D Axial slice of FLAIR MRI before normalization]{{\includegraphics[width=0.3\linewidth,height=0.25\linewidth]{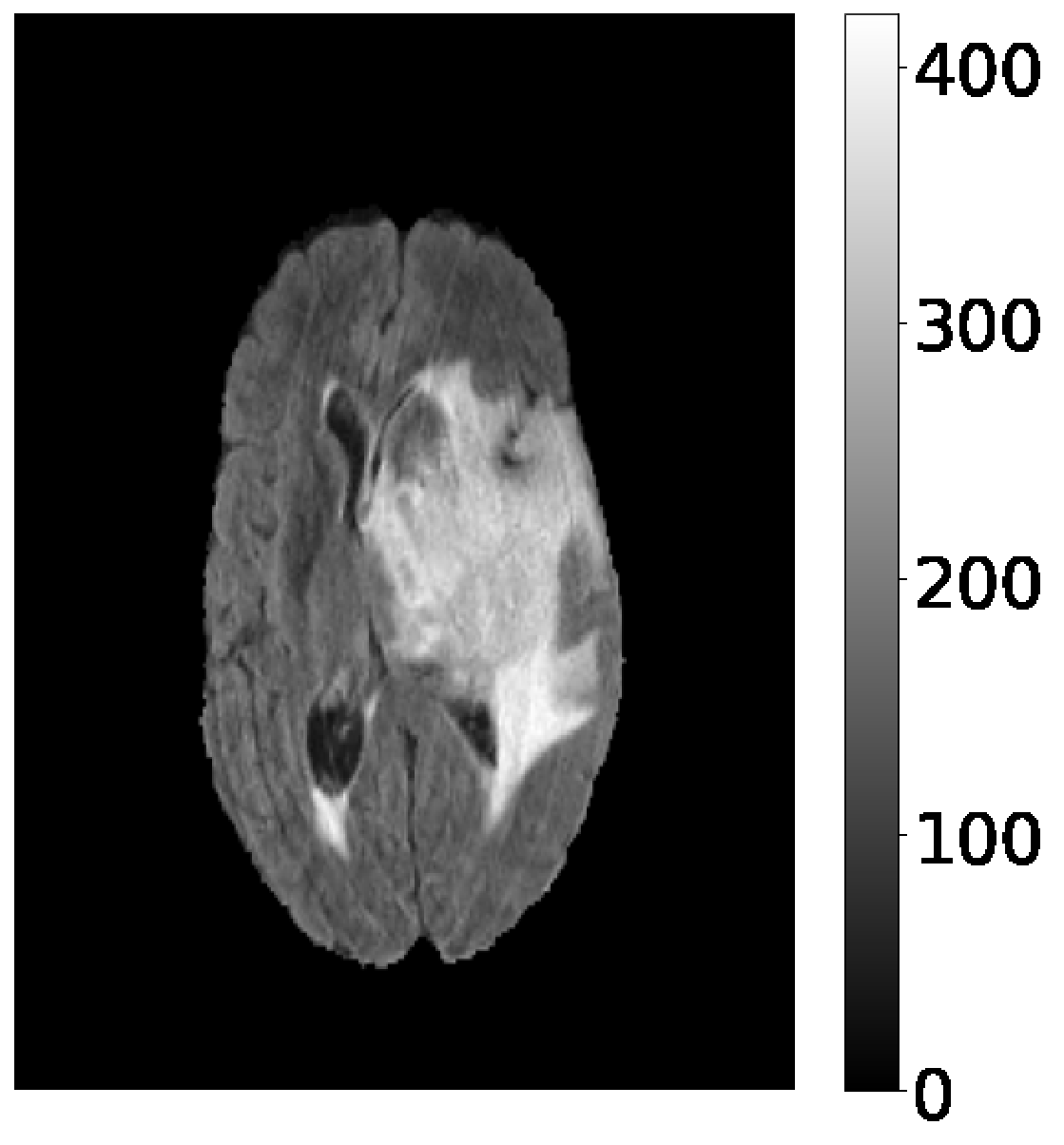} }}
	\quad
	\subfloat[2D Axial slice of T1CE MRI before normalization]{{\includegraphics[width=0.3\linewidth,height=0.25\linewidth]{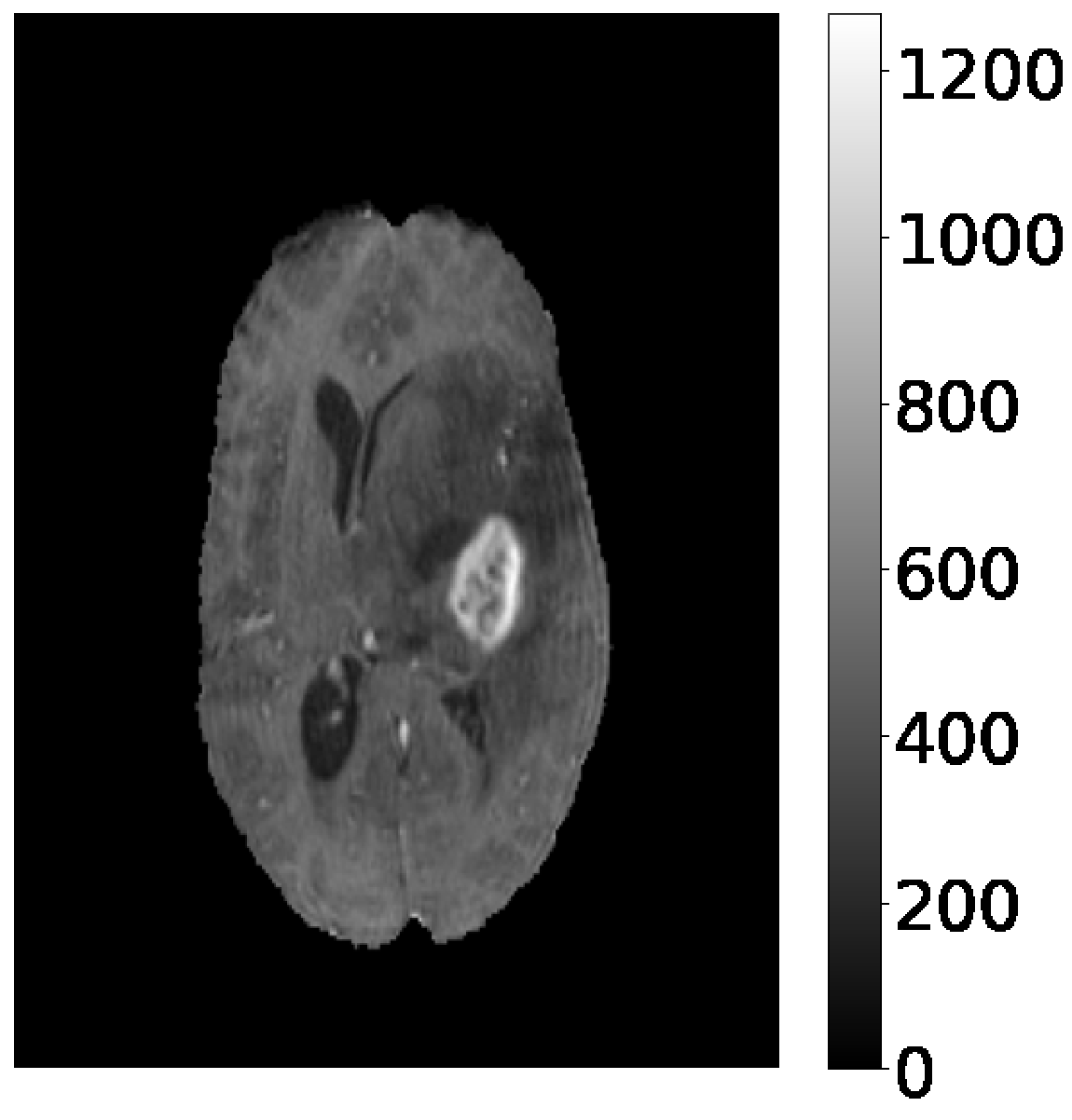} }}
	\quad
	\subfloat[2D Axial slice of T2 MRI before normalization]{{\includegraphics[width=0.3\linewidth,height=0.25\linewidth]{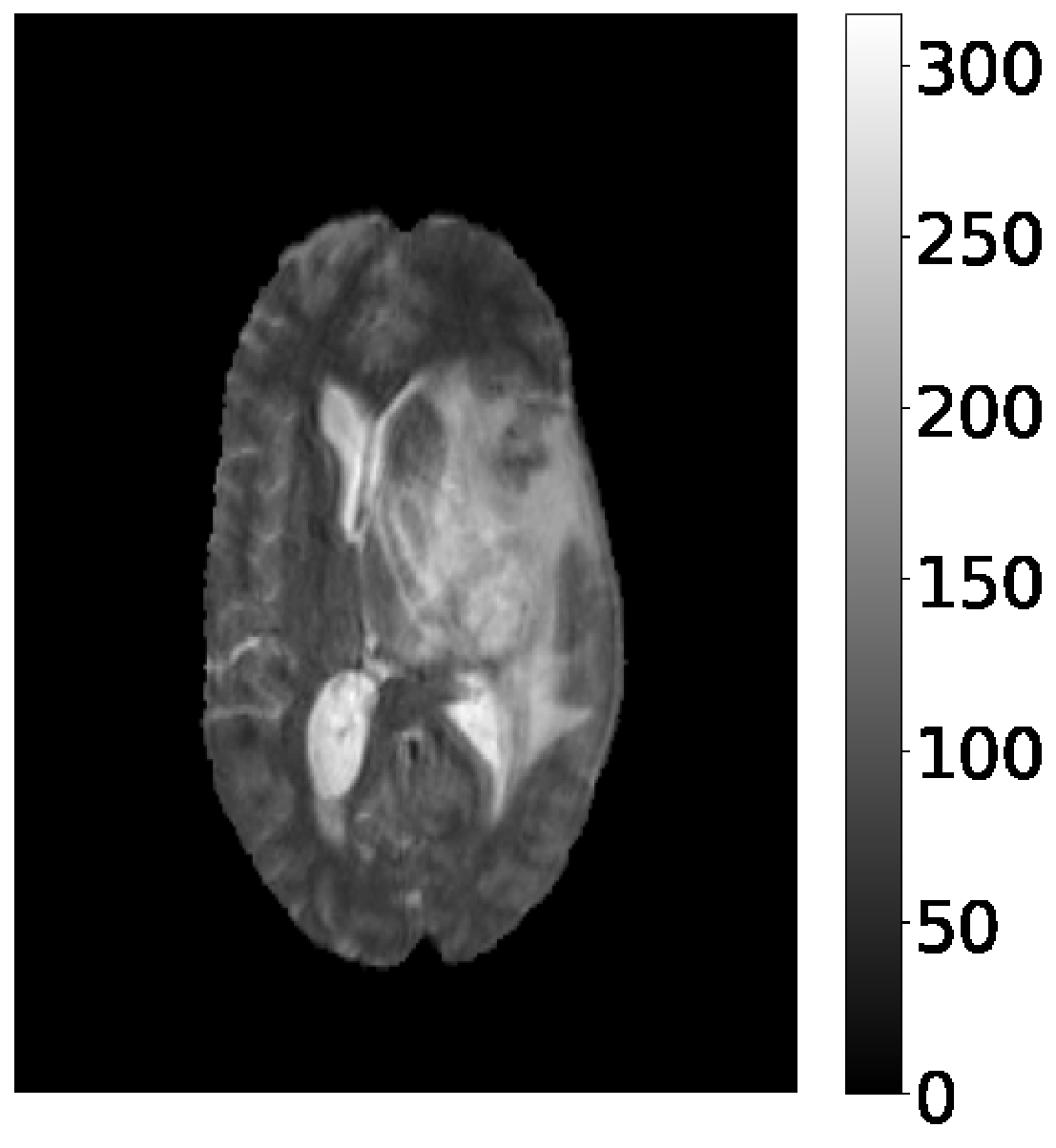} }}
	\quad
	\subfloat[2D Axial slice of FLAIR MRI after normalization]{{\includegraphics[width=0.3\linewidth,height=0.25\linewidth]{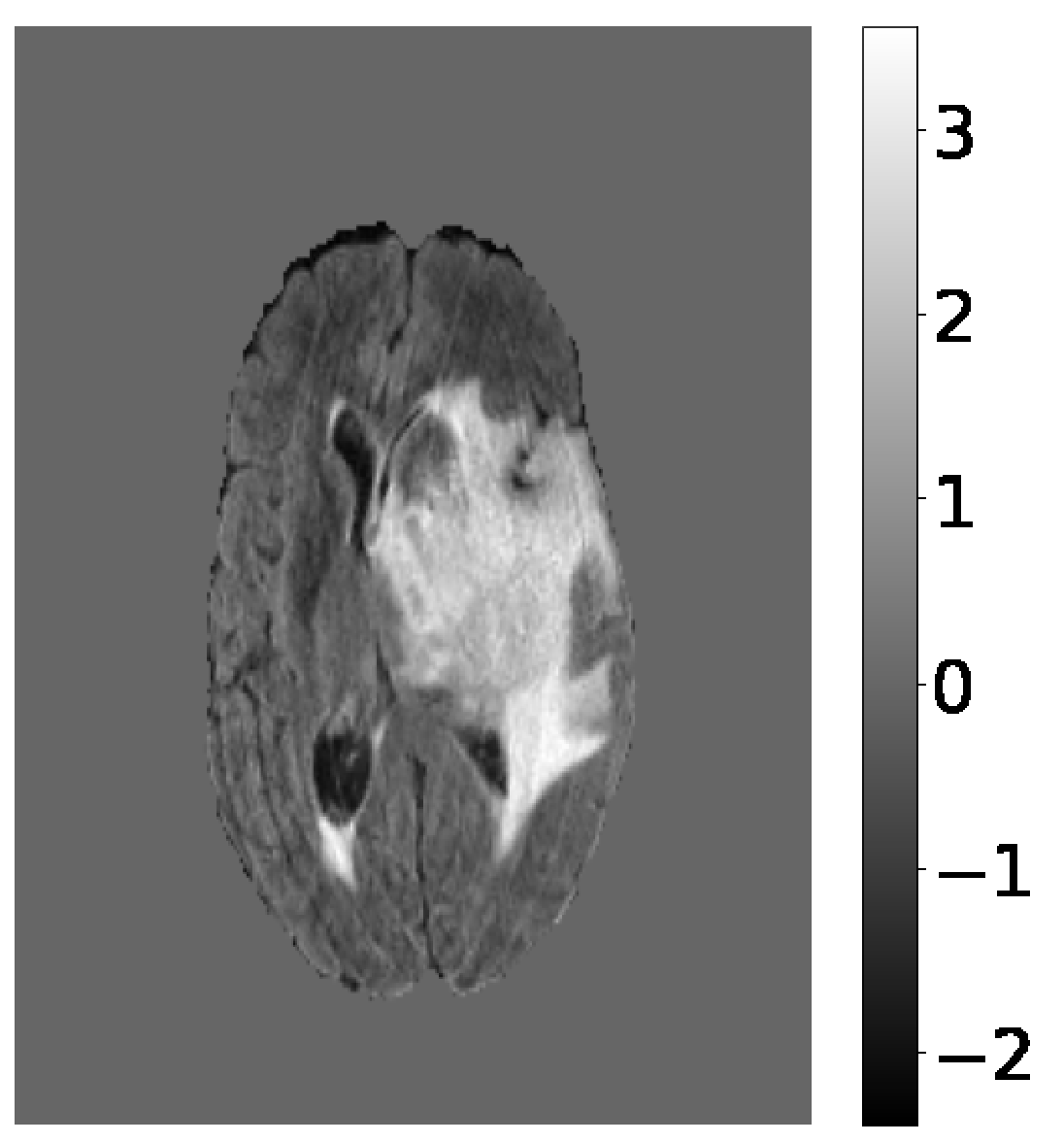} }}
	\quad
	\subfloat[2D Axial slice of T1CE MRI after normalization]{{\includegraphics[width=0.3\linewidth,height=0.25\linewidth]{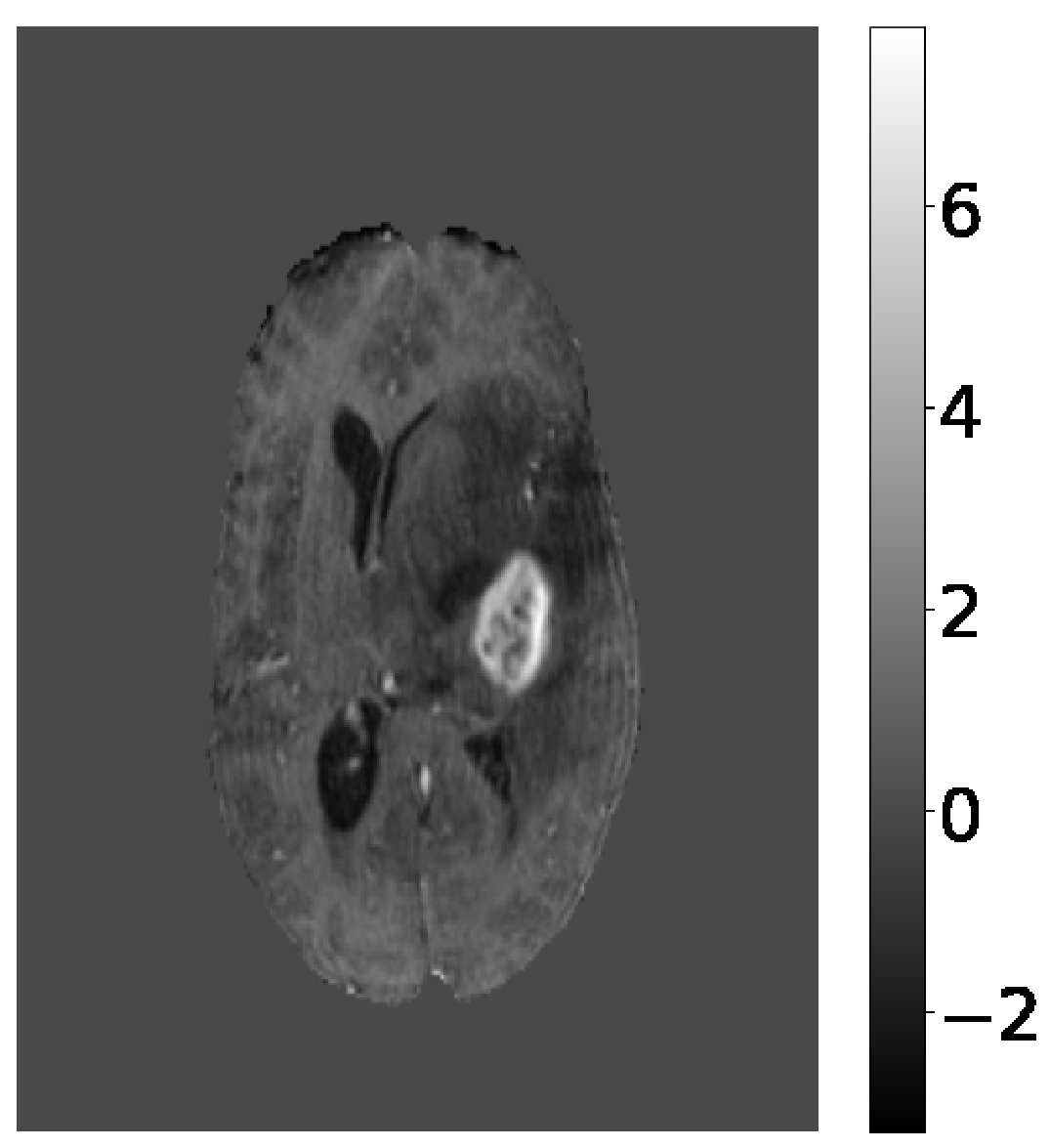} }}
	\quad
	\subfloat[2D Axial slice of T2 MRI after normalization]{{\includegraphics[width=0.3\linewidth,height=0.25\linewidth]{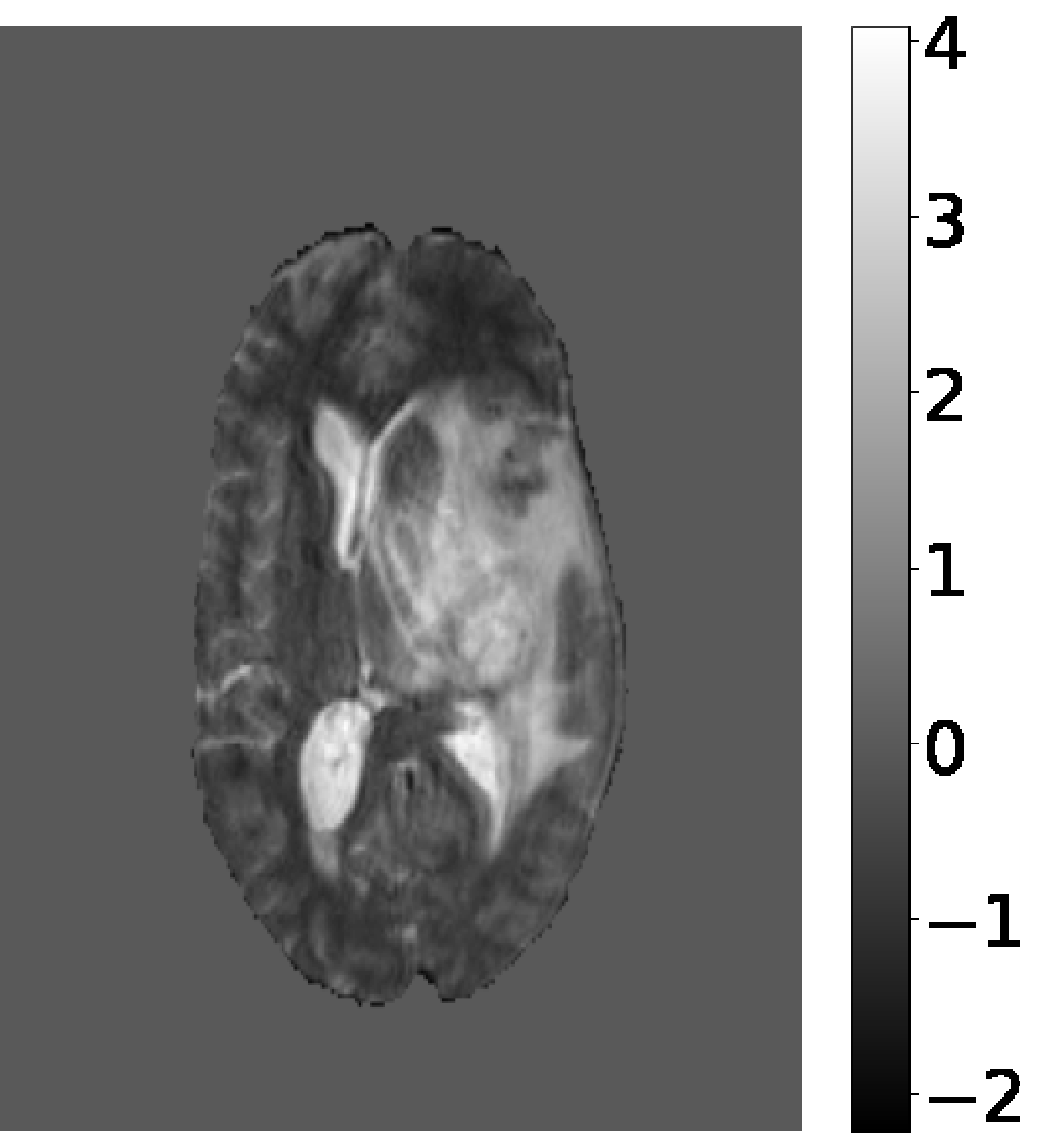} }}
	\caption[Images after performing z-score normalization]{Shows the sample set of 2D axial slice input images (FLAIR, T1ce, T2) before (a-c) and the corresponding MR images (FLAIR, T1ce, T2) after (d-f) Z-score normalization for a patient in BraTS2020 dataset }

\label{fig:mr_input_and_output16}
\end{figure}

The pre-processing steps on the input MR image and target images are pre-requisites before training the deep learning model. Z-score normalization scales the pixel intensity values of the input MR image to have a mean of zero and a standard deviation of one, and is shown in Figure \ref{fig:mr_input_and_output16}. It is a standard technique in medical imaging to account for variations in image acquisition parameters and to ensure consistent pixel intensity values across different images. The z-score normalization is applied only to brain tissue regions in the MR images, the rest remain unchanged.

\begin{figure}[hbt]
\captionsetup[subfigure]{justification=centering}
	\centering
	\subfloat[GroundTruth]{{\includegraphics[width=0.2\linewidth,height=0.2\linewidth]{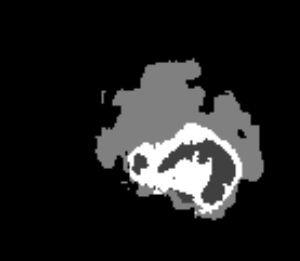} }}
        \quad  
        \subfloat[Edema One-Hot 2D MR Slice]{{\includegraphics[width=0.2\linewidth,height=0.2\linewidth]{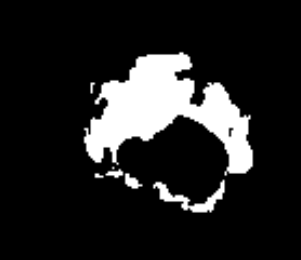} }}
        \quad
	\subfloat[Core One-Hot 2D MR Slice]{{\includegraphics[width=0.2\linewidth,height=0.2\linewidth]{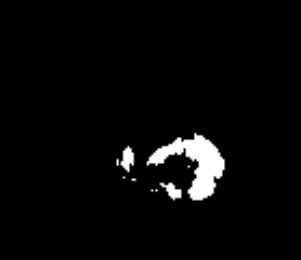} }}
	\quad\\
     \subfloat[Enhancing One-Hot 2D MR Slice]{{\includegraphics[width=0.2\linewidth,height=0.2\linewidth]{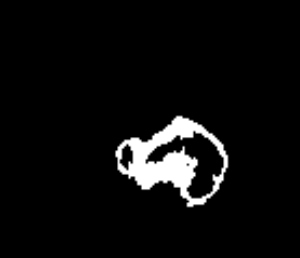} }}
        \quad
        \subfloat[Background One-Hot 2D MR Slice]{\fbox{{\includegraphics[width=0.2\linewidth,height=0.2\linewidth]{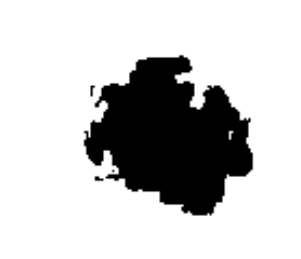} }}}
	
	\caption[One-hot representation of ground truth MR image]{Shows the sample set of 2D MR images with the ground truth (a) and the corresponding one-hot representation of each tumor region (b) - (e).}

\label{mul16}
\end{figure}

 The target images have been converted into one-hot vectors. It encodes the segmentation labels as binary vectors, with each dimension corresponding to a distinct label, Edema, NET/NCR, or ET. The normalized input MR images are then stacked. The ground truth and the corresponding one-hot vectors are shown in Figure \ref{mul16}. These pre-processing steps ensure that the input data is in a suitable format for training the deep learning model.

\subsection{Dynamic Batch Training}

In the proposed work, a sample corresponds to a whole patient's MR stacked input and output images. Assume there are $n$ samples in the dataset. The first epoch in the dynamic batch training is the same as the traditional mini-batch training. However, during this epoch, the image ID and its calculated loss after a forward pass through the deep learning model are stored as a tuple in a list. At the end of the first epoch, this list is sorted in non-increasing order of the loss values.

\begin{figure}[t]
    \centering
    \resizebox{\columnwidth}{!}{
    \begin{tikzpicture}
[align=center, inner sep=0.8mm,thick,
batch/.style={circle,draw=blue!50,fill=blue!20,thick, inner sep=0pt,minimum size=9mm}, 
model/.style={rectangle, draw=orange!50, fill=orange!20, thick, inner sep=5pt, minimum height=1.5cm,minimum width=2cm},
batch_pair/.style={rectangle split, rectangle split horizontal, rectangle split parts=2, draw=red!80, fill=red!20, thick, inner sep=4pt,minimum width=6pt, minimum height=7.2mm},
sorting/.style={rectangle, draw=yellow!80, fill=yellow!40, thick, inner sep=5pt, minimum height=1.5cm,minimum width=1cm},
delta_batch/.style={rectangle split, rectangle split horizontal, rectangle split parts=2, draw=green!70, fill=green!30, thick, inner sep=4pt,minimum width=6pt, minimum height=7.2mm},
every node/.style={transform shape}
]
\draw [decorate,decoration={brace,amplitude=10pt,mirror},xshift=-20pt] (-2.2,4.25) -- (-2.2,-4.25) 
node [black,midway,xshift=-0.8cm,rotate=-90] (n_batches) {$N$ patients};

\draw [gray, densely dashed] (-1.8, 4) -- (0, 0) {};
\draw [gray, densely dashed] (-1.8, 3) -- (0, 0) {};
\draw [gray, densely dashed] (-1.8, 2) -- (0, 0) {};
\draw [gray, densely dashed] (-1.8, 1) -- (0, 0) {};
\draw [gray, densely dashed] (-1.8, 0) -- (0, 0) {};
\draw [gray, densely dashed] (-1.8, -1) -- (0, 0) {};
\draw [gray, densely dashed] (-1.8, -2) -- (0, 0) {};
\draw [gray, densely dashed] (-1.8, -3) -- (0, 0) {};
\draw [gray, densely dashed] (-1.8, -4) -- (0, 0) {};

\node at ( -2.2,4) [batch] {$b_0$};
\node at ( -2.2,3) [batch] {$b_1$};
\node at ( -2.2,2) [batch] {$b_2$};
\node at ( -2.2,1) [batch] {$b_3$};
\node at ( -2.2,0) [batch] {$.$};
\node at ( -2.2,-1) [batch] {$.$};
\node at ( -2.2,-2) [batch] {$.$};
\node at ( -2.2,-3) [batch] {$b_{N-2}$};
\node at ( -2.2,-4) [batch] {$b_{N-1}$};

\node[] at (1.35,1.25) {Forward Pass};
\draw [->] (0.1, 1) -- (2.6, 1) {};
\node at (1.35, 0) [model] (network) {Neural Network};
\draw [<-] (0.1, -1) -- (2.6, -1) {};
\node[] at (1.35,-1.35) {Back Propogation};



\node[thick] at (-1,5.5) {\textbf{Mini-batch training method}};

\draw [thin] (4.5, 5.8) -- (4.5, -5.8) {};

\draw [gray, densely dashed] (6.3, 4) -- (8.6, 0) {};
\draw [gray, densely dashed] (6.3, 3) -- (8.6, 0) {};
\draw [gray, densely dashed] (6.3, 2) -- (8.6, 0) {};
\draw [gray, densely dashed] (6.3, 1) -- (8.6, 0) {};
\draw [gray, densely dashed] (6.3, 0) -- (8.6, 0) {};
\draw [gray, densely dashed] (6.3, -1) -- (8.6, 0) {};
\draw [gray, densely dashed] (6.3, -2) -- (8.6, 0) {};
\draw [gray, densely dashed] (6.3, -3) -- (8.6, 0) {};
\draw [gray, densely dashed] (6.3, -4) -- (8.6, 0) {};

\node at ( 6,4) [batch] {$b_0$};
\node at ( 6,3) [batch] {$b_1$};
\node at ( 6,2) [batch] {$b_2$};
\node at ( 6,1) [batch] {$b_3$};
\node at ( 6,0) [batch] {$.$};
\node at ( 6,-1) [batch] {$.$};
\node at ( 6,-2) [batch] {$.$};
\node at ( 6,-3) [batch] {$b_{N-2}$};
\node at ( 6,-4) [batch] {$b_{N-1}$};

\node[] at (9.98,1.25) {Forward Pass};
\draw [->] (8.7, 1) -- (11.2, 1) {};
\node at (9.98, 0) [model] {Neural Network};
\draw [<-] (8.7, -1) -- (11.2, -1) {};
\node[] at (9.98, -1.35) {Back-Propagation};

\draw [gray, densely dashed] (13.1, 4) -- (11.4, 0) {};
\draw [gray, densely dashed] (13.1, 3) -- (11.4, 0) {};
\draw [gray, densely dashed] (13.1, 2) -- (11.4, 0) {};
\draw [gray, densely dashed] (13.1, 1) -- (11.4, 0) {};
\draw [gray, densely dashed] (13.1, 0) -- (11.4, 0) {};
\draw [gray, densely dashed] (13.1, -1) -- (11.4, 0) {};
\draw [gray, densely dashed] (13.1, -2) -- (11.4, 0) {};
\draw [gray, densely dashed] (13.1, -3) -- (11.4, 0) {};
\draw [gray, densely dashed] (13.1, -4) -- (11.4, 0) {};

\node at ( 13.4,4) [batch_pair] {\nodepart{one} $b_0$ \nodepart{two} $\mathcal{L}_0$};
\node at ( 13.4,3) [batch_pair] {\nodepart{one} $b_1$ \nodepart{two} $\mathcal{L}_1$};
\node at ( 13.4,2) [batch_pair] {\nodepart{one} $b_2$ \nodepart{two} $\mathcal{L}_3$};
\node at ( 13.4,1) [batch_pair] {\nodepart{one} $b_3$ \nodepart{two} $\mathcal{L}_3$};
\node at ( 13.4,0) [batch_pair] {\nodepart{one} $.$ \nodepart{two} $.$};
\node at ( 13.4,-1) [batch_pair] {\nodepart{one} $.$ \nodepart{two} $.$};
\node at ( 13.4,-2) [batch_pair] {\nodepart{one} $.$ \nodepart{two} $.$};
\node at ( 13.4,-3) [batch_pair] {\nodepart{one} $b_{N-2}$ \nodepart{two} $\mathcal{L}_{N-2}$};
\node at ( 13.4,-4) [batch_pair] {\nodepart{one} $b_{N-1}$ \nodepart{two} $\mathcal{L}_{N-1}$};

\draw [-latex, thick] (14, 0) -- (14.8, 0) {};
\node at (16.2, 0) [sorting,align=center] {Sorting\\ mini-batches\\ in desc. order\\ of loss ($\mathcal{L}_i$)};

\draw [-latex, thick] (17.6, 0) -- (18.4, 0) {};


\draw [gray, densely dashed] (20.4, 4) -- (21.9, 0) {};
\draw [gray, densely dashed] (20.4, 3) -- (21.9, 0) {};
\draw [gray, densely dashed] (20.4, 2) -- (21.9, 0) {};
\draw [gray, densely dashed] (20.4, 1) -- (21.9, 0) {};
\draw [gray, densely dashed] (20.4, 0) -- (21.9, 0) {};
\draw [gray, densely dashed] (20.4, -1) -- (21.9, 0) {};

\node at (19.9,4) [delta_batch] {\nodepart{one} $b'_0$ \nodepart{two} $\mathcal{L'}_0$};
\node at (19.9,3) [delta_batch] {\nodepart{one} $b'_1$ \nodepart{two} $\mathcal{L'}_1$};
\node at (19.9,2) [delta_batch] {\nodepart{one} $b'_2$ \nodepart{two} $\mathcal{L'}_2$};
\node at (19.9,1) [delta_batch] {\nodepart{one} $.$ \nodepart{two} $.$};
\node at (19.9,0) [delta_batch] {\nodepart{one} $b'_{\delta N - 2}$ \nodepart{two} $\mathcal{L'}_{\delta N - 2}$};
\node at (19.9,-1) [delta_batch] {\nodepart{one} $b'_{\delta N - 1}$ \nodepart{two} $\mathcal{L'}_{\delta N - 1}$};
\node at (19.9,-2) [batch_pair, opacity=0.5] {\nodepart{one} $.$ \nodepart{two} $.$};
\node at (19.9,-3) [batch_pair, opacity=0.5] {\nodepart{one} $b'_{N-2}$ \nodepart{two} $\mathcal{L'}_{N-2}$};
\node at (19.9,-4) [batch_pair, opacity=0.5] {\nodepart{one} $b'_{N-1}$ \nodepart{two} $\mathcal{L'}_{N-1}$};

\node[] at (23.26,1.25) {Forward Pass};
\draw [->] (21.9, 1) -- (24.5, 1) {};
\node at (23.26, 0) [model] {Neural Network};
\draw [<-] (21.9, -1) -- (24.5, -1) {};
\node[] at (23.26, -1.35) {Back Propagation};

\draw[] (24.7, 0) -- (24.9, 0) {};
\draw[] (24.9, 0) -- (24.9, -5) {};
\draw[] (24.9, -5) -- (16.2, -5) {};
\draw[->, -latex] (16.2, -5) -- (16.2, -1.25) {};


\node[] at (20.6, -5.5) {\Large Repeat (1/$\delta$) times};

\node[thick] at (13.9,5.5) {\textbf{Proposed method}};


\end{tikzpicture}
}
    \caption{An overview of the existing mini-batch training method \cite{ruder2016overview} (left) and the proposed method (right)}
    \label{fig: workflow}
\end{figure}

After the first epoch, dynamic batch training begins. A hyper-parameter $\delta$ is introduced, which can take values between $(0, 1]$. Based on the value of $\delta$, the top ($\delta \times n$) samples, corresponding to the highest loss values, are selected for training. This selection and training process is repeated $1/\delta$ times. Therefore, during each epoch after the first, the model is trained on samples with the highest loss values, effectively focusing on the more challenging cases. This ensures that the model sees all data points in the first epoch and then continues to focus on the hardest samples in subsequent epochs. The main difference between this and traditional mini-batch training is that in the proposed method, after the first epoch, the model prioritizes poorly performing images, whereas traditional methods treat all samples equally. The key difference is that, in the proposed method, the model focuses more on difficult samples with higher loss values, potentially skipping easier samples that consistently yield low loss.

Figure \ref{fig: workflow} illustrates the traditional mini-batch training on the left and the proposed dynamic batch training on the right. To reduce the risk of overfitting and to stabilize the learning process, dynamic batch training is alternated with normal training across epochs. Specifically, dynamic batch training occurs on even-numbered epochs, while normal training occurs on odd-numbered epochs. The complete algorithm of the proposed dynamic batch training approach is described in Algorithm \ref{alg:dybat}.

\begin{algorithm}[tbh]
\DontPrintSemicolon
  \KwInput{Number of epochs $E$, Model $\mathcal{W}$ }
  \KwOutput{Trained weights $\mathcal{W}$}
  \KwData{$\mathcal{D}_{T}$ = Train Dataset $(input, target)^{N}_{i = 0}$ where N are the total batches}
  \For{$e = 0, 1, \ldots, (E-1)$}{
    \tcc{Training $N$ patients, $input_i$ will have 64 MR Images}
    \For{$(input_i, target_i) \in \mathcal{D}_T$}{
        \tcc{Forward Pass}
        $pred_i = \mathcal{W}(input_i)$ \\
        \tcc{Calculate Train Loss}
        $\mathcal{L} \gets (target_i, pred_i)$ \\
        \tcc{Back propagate loss on $\mathcal{W}$}
        $\mathcal{W} \gets \mathcal{L}$ \\
    }
  }
  \Return{$\mathcal{W}$}
\caption{Traditional training approach}
\label{alg:standard}
\end{algorithm}

\begin{algorithm}[H]
\DontPrintSemicolon
  \KwInput{Hyper-parameter $\delta$, Epochs $E$, Network $\mathcal{W}$}
  \KwOutput{Trained weights $\mathcal{W}$}
  \KwData{$\mathcal{D}_{T}$ = Train Dataset $(input, target)^{N}_{i = 0}$ where N are the total batches}
  \tt{List = []} \\
  \For{$e = 0, 1, \ldots, (E-1) / 2$}{
    \tcc{Regular training step}
  \For{$(input_i, target_i) \in \mathcal{D}^T$}{
        $p = \mathcal{W}(input_i)$ \\
       $\mathcal{L} \gets (target_i, p)$ \\
        count\_i $\gets$ count\_i + 1 \\
        $\tt{List[i]} \gets (patient_{id}, \mathcal{L}, count_i)$ \\
        $\mathcal{W} \gets \mathcal{L}$
    }
     \tcc{Dynamic batch training with $\delta \times N$ mini-batches}
  \For{$z = 0, 1, \ldots, (\frac{1}{\delta})$}{
  \tcc{Sort List in descending order of $\mathcal{L} per batch$}
    Sort(\tt{List}) \\
    \tcc{Train on first $\delta \times N$ batches}
    $\mathcal{D} \gets \{(input_i, target_i)\}_{i=0}^{\delta \times N}$ \\
    \For{$(input_i, target_i) \in \mathcal{D}$}{
        $pred_i = \mathcal{W}(input_i)$ \\
        $\mathcal{L} \gets (target_i, pred_i)$ \\
        count\_i $\gets$ count\_i + 1 \\
        $\tt{List[i]} \gets (patient_{id}, \mathcal{L}, count_i)$ \\
        $\mathcal{W} \gets \mathcal{L}$\\
    }
  }
  }
  \Return{$\mathcal{W}$}
\caption{Proposed dynamic batch training approach}
\label{alg:dybat}
\end{algorithm}

\section{Experimental Results}
\label{chap: results}

The experimental section details the dataset, evaluation metrics, loss functions used for training the models, the implementation details, the ablation study, the quantitative study, and the comparison study.

\subsection{Dataset}

The BraTS2020 \cite{bakas2017segmentation, bakas2017segmentationgbm} dataset is used for all the experiments and studies carried out in this work. The dataset contains pre-operative scans of high-grade and low-grade gliomas(HGG). It includes data from The Cancer Imaging Archive (TCIA) glioma collections and $19$ institutes worldwide and has been labeled by a group of expert neuroradiologists. The dataset has 369 patients for training and 125 patients for validation. Each of the patients has four MRI sequences: T1-weighted (T1), T1-weighted contrast-enhanced (T1CE), T2-weighted (T2), and Fluid Attenuated Inversion Recovery (FLAIR). The training data additionally contains a ground-truth MR image. The T1-weighted image generally contains no valuable information for the brain tumor segmentation tasks and has not been used. The BraTS2020 dataset contains patients from both previous BraTS challenges, and it can be considered a super-set of BraTS2018 and BraTS2019 datasets. 

\begin{figure}[htb]
\captionsetup[subfigure]{justification=centering}
	\centering
	\subfloat[FLAIR Image]{{\includegraphics[width=0.21\linewidth]{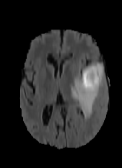} }}
	\quad
	\subfloat[T1CE Image]{{\includegraphics[width=0.21\linewidth]{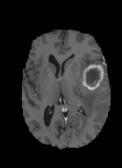} }}
	\quad
	\subfloat[T2 Image]{{\includegraphics[width=0.21\linewidth]{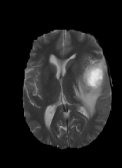} }}
	\quad
	\subfloat[Ground Truth]{{\includegraphics[width=0.21\linewidth]{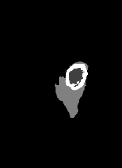} }}
	
	\caption[Sample Input and Output MR Image]{Shows a sample 2D axial input MR Images (a-c) and the corresponding ground truth (d). In the ground truth, the white region corresponds to the enhancing tumor (ET), the dark grey region corresponds to the necrosis and non-enhancing tumor (NCR/NET), and the light grey area represents the edema region (ED) \cite{sahayam2022brain}.}
 \label{mr_input_and_output1}
\end{figure}

The dataset's MR images are preprocessed so that the skulls and neck regions are removed. The images are aligned and co-registered to correct any head tilt or mismatches between MR modalities. Each $1$ voxel in the image corresponds to $1 mm^{3}$ of brain tissue. Figure \ref{mr_input_and_output1} illustrates a 2D axial input MR Images sample and the corresponding ground truth. The input images (a-c) show the different MR modalities: FLAIR, T1CE, and T2 images. The ground truth consists of peritumoral edema (ED) marked in light grey given by an intensity value of $2$,  enhancing tumor (ET) represented as a white region with $4$ as an intensity value, and the non-enhancing tumor (NET) and necrotic core region (NCR) as dark grey with an intensity value of $1$. The models are evaluated as enhancing tumor (ET), tumor core (TC), and whole tumor (WT). The tumor core corresponds to NET/NCR and the enhancing tumor region. The whole tumor corresponds to all three tumor regions put together. The research aims to learn the ED, NET/NCR, and ET regions.

\subsection{Evaluation Metrics}
Performance metrics are used to evaluate the effectiveness of the predicted segmentation when compared with the ground truth. The metrics used for measuring the segmentation performance are the dice score and the bi-direction Hausdorff95 distance. 

The dice score \cite{atzeni2022deep} is a standard evaluation metric used to quantify the performance of a segmentation algorithm. It measures the overlap between the predicted segmentation and the ground truth segmentation by calculating the ratio of twice the number of true positive (TP) voxels to the sum of TP, false positive (FP), and false negative (FN) voxels. The TP voxels are the number of correctly segmented voxels, while the FP voxels are the number of voxels that have been incorrectly classified as foreground tissue. The FN voxels are the number of voxels that belong to the background but have been incorrectly predicted as foreground. A Dice score of 1 indicates perfect segmentation, while a score of 0 indicates no overlap between the predicted and ground truth segmentation. The dice score is widely used because it considers both false positives and false negatives and is sensitive to minor errors in the segmentation. The dice score is given by Equation \ref{dice}.

\begin{equation}
    \textrm{Dice Score =} \frac{2*TP}{(TP + FP) + (TP + FN)}
    \label{dice}
\end{equation}

Hausdorff distance \cite{van2022between} is a metric used in image segmentation tasks to quantify the difference between two sets of points. The Hausdorff distance is the maximum distance between a point in one set and its nearest point in the other. It helps to identify whether the predicted segmentation mask and the ground truth segmentation mask are identical across the edge regions \cite{huttenlocher1993comparing}. Higher values of Hausdorff distance indicate a poor model performance between the predicted and ground truth. A Hausdorff distance of $1$ indicates a perfect segmentation performance. The general one-sided HD from X to Y and Y to X are given by Equations \ref{eq4} and \ref{eq5}.
\begin{equation}
        \textrm{$\hat{H}(X,Y)$ = } max(min_{x \in X}d(x,Y))
        \label{eq4}
\end{equation}
\begin{equation}
    \textrm{$\hat{H}(Y,X)$ = } max(min_{y \in Y}d(X,y))\
    \label{eq5}
\end{equation}

The one-sided HD distance is not a metric. It fails the symmetry property of a metric. However, the bidirectional HD is a metric. The bidirectional HD between these two sets is defined in Equation \ref{eq6}.
\begin{equation}
    \textrm{$H(X,Y)$ = }max(\hat{H}(X,Y), \hat{H}(X,Y))
    \label{eq6}
\end{equation}
Hausdorff distance is sensitive to outliers and noise. So, 95\% of the Hausdorff value is used with a 5\% threshold for error.

\subsection{Loss Function}
The BraTS2020 dataset suffers from a high-class imbalance. Focal loss is a function that handles class imbalance in binary classification tasks. Focal loss assigns a low weight to the class in most instances. The weight is managed by a parameter $\alpha_{f}$. $\gamma$ is a modulating factor that reduces the weight of easy examples while increasing the weight of hard examples. A $\gamma$ value of $2$ is set for training the different deep-learning models in the proposed work \cite{focal_tversky_loss}.
\begin{equation}
   \textrm{Focal Loss $L(y,\hat{y})$ = }-\alpha_{f} y(1-\hat{y})^{\gamma}log(\hat{y})-(1-y)\hat{y}^{\gamma}log(1-\hat{y})
\end{equation}
The value of alpha varies depending on the target class as follows,
 where \[
    \alpha_{f}= 
\begin{cases}
   0.8,& \text{if } \textrm{target} \in \{\textrm{NET/NCR, ED, ET}\}\\
    0.2,              & \textrm{Background}
\end{cases}
\]
There are a few under-represented samples in the BraTS2020 dataset that do not contain a specific tumor class in the whole MR image. The BraTS online evaluation engine assigns a high penalty of $0$ to dice and $373.12866$ for Hausdorff95 metric to such cases. The focal loss function for such tumor region is calculated by setting the $\alpha_{f}$ value to $0$. The focal loss will be reduced to Equation \ref{FPFL} as follows,
\begin{equation}
   \textrm{ False Positive Focal Loss $L(y,\hat{y})$ = }-(1-y)\hat{y}^{\gamma}log(1-\hat{y})
   \label{FPFL}
\end{equation}
The above loss is added with the loss for the other tumor regions as follows,
\begin{equation}
   \textrm{ Hybrid Focal Loss $L(y,\hat{y})$ = } \textrm{Focal Loss} + C \times \textrm{False Positive Focal Loss}
\end{equation}
where \textrm{$C$ is a constant.}

Additionally, instead of the false positive focal loss, the mean false positive loss has been proposed in Equation \ref{MFPL} as follows,
\begin{equation}
   \textrm{ Mean False Positive Loss $L(y,\hat{y})$ = }\frac{1}{n} \times(1-y) \times \hat{y}
   \label{MFPL}
\end{equation}
where \textrm{n is the number of voxels in the image.}
The above loss is added with the loss for the other tumor regions as follows,
\begin{equation}
   \textrm{ Mean False Positive Focal Loss $L(y,\hat{y})$ = } \textrm{Focal Loss} +  \textrm{Mean False Positive Loss}
\end{equation}

As the penalty for false positives is maximum in the absence of tumor regions, the hybrid focal loss, and the mean false positive focal loss have been calculated only for instances where the entire tumor region is absent for a patient in a training batch. Also, if the tumor region is small and appears only in a few slices, there is a chance that some of the few slices may not contain the tumor region at all. In such cases, the two proposed focal losses will be calculated for that specific batch. For example, if there are $155$ 2D slices for a patient and assume the tumor occupies only $10$ frames between frames $90$ and $100$. If the batch size is $64$, then the patient information will be split into 3 batches. The first two batches will contain $64$ 2D slices, and the last will contain $27$ 2D slices. The tumor region loss for the first and the last batch will be calculated using either one of the proposed false positive focal losses and the tumor region loss for the second batch will be the regular focal loss. The total loss will be the sum of the three loss function values.

\subsection{Implementation Details}
The batch size for training is $64$, and the Adam optimizer with a learning rate of $0.01$ is used for all models, with all hyperparameters kept the same for a fair comparison. The models are trained using Keras \cite{chollet2015keras}, and TensorFlow \cite{tensorflow2015-whitepaper}. The segmentation models are designed using the Keras U-Net collection \cite{keras-unet-collection}.

The axial view of each of the patients has been used in the training process. None of the frames are dropped. For a patient, a 240 x 240 2D axial slice is stacked so that the input image would be of dimension 240 x 240 x 3. The stacked frames are the corresponding frames from FLAIR, T1CE, and T2 MR images for a patient. A patient will have 155 such stacked 2D axial input frames. After stacking, the input image would be 155 x 240 x 240 x 3. The Z-score normalization of the inputs, edge extraction from the ground truth images, and the one-hot representation of the ground truth and the edges have been performed. The one-hot images are stacked on top of each other. If the training consisted of only the tumor regions, the stacked one-hot representation will have a dimension of 155 x 240 x 240 x 4, with the last dimension corresponding to the background, NET/NCR, edema, and enhancing, respectively, with 155 slices per patient. All the models used ReLU as an activation function for a fair comparison and softmax activation in the last layer to get the final tumor prediction.  Adam \cite{kingma2014adam} has been used as an optimizer with an initial learning rate $\alpha$ of 0.01. The models have been trained for a total of $50$ epochs. The model weights are saved for epochs with the highest average dice score value. The models have been trained on NVIDIA A100 with 40GB RAM.

\subsection{Ablation Study}
An ablation study is used to evaluate the contribution of individual components in a system. It involves systematically removing one or more components of the system to determine their impact on the overall performance of the system. Ablation study is a helpful tool for understanding the workings of deep learning models. It can help to identify the most essential features or components of a model and to assess the performance of the model to changes in those components. In addition, ablation studies can aid in identifying redundant or less practical components, leading to more efficient models. There are two ablation studies performed in the proposed work, namely, ablation to find $\delta$ and ablation to find optimum weight $C$ for the hybrid focal loss.
\subsubsection{Ablation on Delta}
Table \ref{tab: ablation_train_delta} displays the training results of the Hybrid MR U-Net model for different values of hyperparameter $\delta$. The $\delta$ value controls the number of hard samples selected during the dynamic batch training process. The model's performance has been evaluated using the dice score and Hausdorff95 distance metric for three different tumor regions, namely, ET, WT, and TC, on the BraTS2020 dataset. The best-performing model is highlighted in bold. Among the different $\delta$ values, the highest overall dice score of $0.765$ and the lowest Hausdorff95 distance metric for ET of $26.19$ has been achieved for the model trained with a delta equal to $1$. However, an interesting observation is the performance of the model trained with a $\delta$ equal to $0.2$, which achieved a high dice score of $0.750$ for ET and a low Hausdorff95 distance metric for WT $(5.15)$ and TC $(7.24)$. It implies that using dynamic batching with a lower $\delta$ value can lead to better performance in detecting smaller and more complex tumor regions.
\begin{table}[htb]
\centering
\caption{Shows the ablation on the various values of the hyperparameter delta on the training data.}
\begin{tabular}{ccrrrrrr}
\hline
\multicolumn{1}{l}{\multirow{2}{*}{\textbf{Model}}} & \multicolumn{1}{l}{\multirow{2}{*}{\textbf{Delta}}} & \multicolumn{3}{c}{\textbf{Dice Score $\uparrow$}}                                  & \multicolumn{3}{c}{\textbf{Hausdorff95 $\downarrow$}}                                 \\ \cline{3-8} 
\multicolumn{1}{l}{}                                & \multicolumn{1}{l}{}                                & \multicolumn{1}{l}{ET} & \multicolumn{1}{l}{WT} & \multicolumn{1}{l}{TC} & \multicolumn{1}{l}{ET} & \multicolumn{1}{l}{WT} & \multicolumn{1}{l}{TC} \\ \hline
\multirow{3}{*}{Hybrid MR U-Net}                    & 0.2                                                 &  0.750                  & \textbf{0.906}                  & \textbf{0.831}                  & 31.73                  & \textbf{5.15}                   & \textbf{7.24}                   \\
                                                    & 0.5                                                 & 0.747                  & 0.891                 & 0.826                  & 33.91                  & 6.03                   & 8.98                   \\
                                                    & 1   &                                                \textbf{0.765}  &  \textbf{0.906} &  0.775 &  \textbf{26.19}  &  8.07 &   11.07                 \\ \hline
\end{tabular}
\label{tab: ablation_train_delta}
\end{table}

Table \ref{tab: ablation_valid_delta} shows the results of an ablation study on a Hybrid MR U-Net model with different values of the hyperparameter $\delta$ on validation data.

The results show that when that $\delta$ is set to $1$, the model achieves the highest dice scores of $0.691$ and $0.876$ for ET and WT, respectively. However, it has a higher Hausdorff distance value of $34.43$ for ET, indicating more outliers in the segmentation results. But when $\delta$ is set to 0.2, the model achieves the highest dice score of $0.706$ for TC and the lowest Hausdorff distance values of $9.93$ and $16.86$ for WT and TC, respectively. There is a significant improvement in the Hausdorff95 score when training with $\delta$ of $0.2$. Based on these results, the $\delta$ of $0.2$ is a better choice for selecting hard samples as it produces good Hausdorff95 segmentation results compared to using $\delta$ = $1$, despite a slightly lower dice score for ET and WT. For the remaining experiments, $\delta$ of $0.2$ has been used.

\begin{table}[htb]
\centering
\caption{Shows the ablation on the various values of the hyperparameter delta on the validation data.}
\begin{tabular}{ccrrrrrr}
\hline
\multicolumn{1}{l}{\multirow{2}{*}{\textbf{Model}}} & \multicolumn{1}{l}{\multirow{2}{*}{\textbf{Delta}}} & \multicolumn{3}{c}{\textbf{Dice Score $\uparrow$}}                                  & \multicolumn{3}{c}{\textbf{Hausdorff95 $\downarrow$}}                                 \\ \cline{3-8} 
\multicolumn{1}{l}{}                                & \multicolumn{1}{l}{}                                & \multicolumn{1}{l}{ET} & \multicolumn{1}{l}{WT} & \multicolumn{1}{l}{TC} & \multicolumn{1}{l}{ET} & \multicolumn{1}{l}{WT} & \multicolumn{1}{l}{TC} \\ \hline
\multirow{3}{*}{Hybrid MR U-Net}                    & 0.2                                                 & 0.664                  & 0.875                  & \textbf{0.706}                  & 41.62                  & \textbf{9.93}                   & \textbf{16.86}                  \\
                                                    & 0.5                                                 & 0.684                  & 0.860                  & 0.686                  & 41.71                  & 10.44                   & 19.92                  \\
                                                    & 1                                                   & \textbf{0.691}                              & \textbf{0.876}                     &  0.644                              &   \textbf{34.43}               &  14.58                     &  32.93                   \\ \hline
\end{tabular}
\label{tab: ablation_valid_delta}
\end{table}

\subsubsection{Ablation to Find C}
Table \ref{tab: ablation_train_c} shows the results of an ablation study on the various values of false positive weight $C$ for the hybrid focal loss function on the training data. The table reports the dice score and Hausdorff95 values for the Enhancing Tumor (ET), Whole Tumor (WT), and Tumor Core (TC) regions. The Hybrid MR U-Net model has been trained with different values of $C$, namely $5$, $10$, $15$, $20$, and $30$. The results show that the model achieved the highest dice score and lowest Hausdorff95 value for $C$ $=$ $10$. Specifically, the model achieved a dice score of $0.768$, $0.911$, and $0.845$ for ET, WT, and TC regions, respectively. Moreover, the model achieved a Hausdorff95 value of $26.39$, $4.30$, and $4.26$ for ET, WT, and TC regions, respectively.

As the results suggest, setting $C$ to $10$ results in the best performance for the Hybrid MR U-Net model. The model achieves a high dice score and low Hausdorff95 value, indicating that it accurately identifies the segmentation boundaries of the tumor regions.
\begin{table}[htb]
\centering
\caption{Shows the ablation on the various values of C for the hybrid focal loss on the training data.}
\begin{tabular}{cccccccc}
\hline
\multicolumn{1}{l}{\multirow{2}{*}{\textbf{Model}}} & \multicolumn{1}{l}{\multirow{2}{*}{\textbf{C}}} & \multicolumn{3}{c}{\textbf{Dice Score $\uparrow$}}                                  & \multicolumn{3}{c}{\textbf{Hausdorff95 $\downarrow$}}                                 \\ \cline{3-8} 
\multicolumn{1}{l}{}                                & \multicolumn{1}{l}{}                            & \multicolumn{1}{l}{ET} & \multicolumn{1}{l}{WT} & \multicolumn{1}{l}{TC} & \multicolumn{1}{l}{ET} & \multicolumn{1}{l}{WT} & \multicolumn{1}{l}{TC} \\ \hline
\multirow{5}{*}{Hybrid MR U-Net}                    & 5                                               & 0.748                  & 0.894                  & 0.817                  & 31.71                  & 5.19                   & 6.37                   \\
                                                    & 10                                              & \textbf{0.768}                  & \textbf{0.911}                  & \textbf{0.845}                  & \textbf{26.39}                  & \textbf{4.30}                   & \textbf{4.26}                   \\
                                                    & 15                                              & 0.745                  & 0.903                  & 0.835                  & 32.83                  & 9.38                   & 6.45                   \\
                                                    & 20                                              & 0.758                  & 0.902                  & 0.830                  & 28.44                  & 4.55                   & 7.06                   \\
                                                    & 30                                              & 0.738                  & 0.912                  & 0.833                  & 31.54                  & 5.23                   & 5.55                   \\ \hline
\end{tabular}
\label{tab: ablation_train_c}
\end{table}

Table \ref{tab: ablation_valid_c} summarizes the ablation study results on the validation dataset. The table shows the performance of the Hybrid MR U-Net with different values of $C$ ranging from $5$ to $30$. The results indicate that $C$=10 performs significantly better in the HD95 metric in TC. Additionally, while the Dice score for the $C$=10 cases is slightly lower than the highest values obtained with $C$=20 and $C$=30, the differences are marginal and can be ignored. Based on these results, the usage of $C$=10 has been used in the remaining experiments in the study.

\begin{table}[htb]
\centering
\caption{Shows the ablation on the various values of C for the hybrid focal loss on the validation data.}
\begin{tabular}{cccccccc}
\hline
\multicolumn{1}{l}{\multirow{2}{*}{\textbf{Model}}} & \multicolumn{1}{l}{\multirow{2}{*}{\textbf{C}}} & \multicolumn{3}{c}{\textbf{Dice Score $\uparrow$}}                                  & \multicolumn{3}{c}{\textbf{Hausdorff95 $\downarrow$}}                                 \\ \cline{3-8} 
\multicolumn{1}{l}{}                                & \multicolumn{1}{l}{}                            & \multicolumn{1}{l}{ET} & \multicolumn{1}{l}{WT} & \multicolumn{1}{l}{TC} & \multicolumn{1}{l}{ET} & \multicolumn{1}{l}{WT} & \multicolumn{1}{l}{TC} \\ \hline
\multirow{5}{*}{Hybrid MR U-Net}                    & 5                                               & 0.659                  & 0.857                  & 0.682                  & \textbf{39.14}                  & 11.46                  & 17.16                  \\
                                                    & 10                                              & 0.674                  & 0.874                  & \textbf{0.710}                  & 41.62                  & 8.81                   & \textbf{12.81}                  \\
                                                    & 15                                              & 0.641                  & 0.864                  & 0.694                  & 54.70                  & 16.58                  & 15.16                  \\
                                                    & 20                                              & \textbf{0.677}                  & \textbf{0.878}                  & 0.685                  & 46.41                  & \textbf{7.72}                   & 24.48                  \\
                                                    & 30                                              & 0.674                  &\textbf{0.878}                  & 0.704                  & 42.93                  & 11.89                  & 16.47                  \\ \hline
\end{tabular}
\label{tab: ablation_valid_c}
\end{table}

\subsection{Quantitative Analysis}
\begin{figure}[ht]
\centering
	\includegraphics[width=0.8\linewidth]{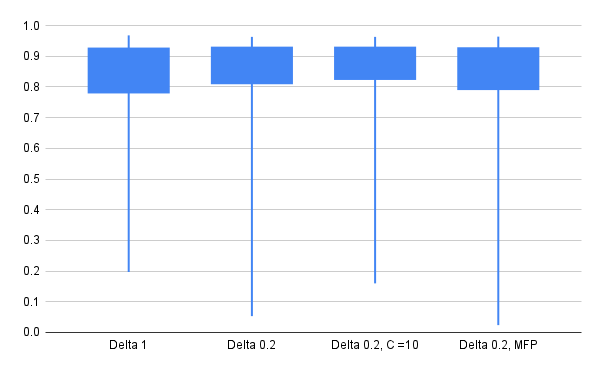}
	\caption{Shows a box-plot on the BraTS2020 training dataset.}
	\label{fig: boxplot}
\end{figure}

\begin{figure}[!ht]
\centering
	\includegraphics[width=0.8\linewidth]{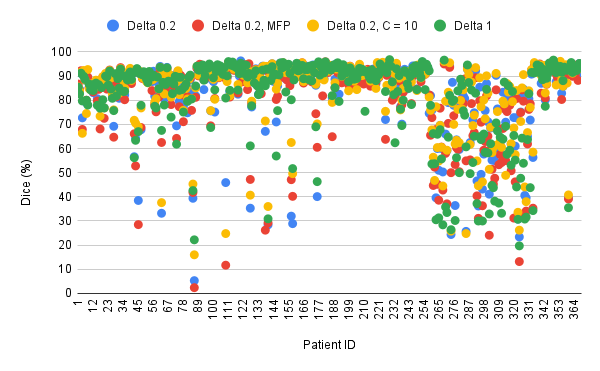}
	\caption{Shows a scatterplot with patient ID vs. average dice score on the BraTS2020 training dataset.}
	\label{fig: pid2dice}
\end{figure}

A box plot is a statistical graph that displays the distribution of a data set by visually representing the minimum, maximum, median, and quartiles. It helps identify outliers. The box in the box plot represents the inter-quartile range (IQR), which is the range between the $25^{th}$ and $75^{th}$ percentile of the data. Figure \ref{fig: boxplot} shows the boxplot for hybrid MR-U-Net trained with $\delta$ = 1, $\delta$ = 0.2 with the regular focal loss, weighted focal loss with $C$ = 10, and mean false positive focal loss. The tails of each box show the minimum and maximum average dice scores obtained by each method. It can be observed that $\delta$ = 0.2 with a weighted focal loss with $C$ = 10 has the shortest IQR with the data skewed towards the higher dice score side. It shows the improved segmentation capabilities of the weighted focal loss with $C$ = 10. The regular training of hybrid MR-U-Net had the shortest minimum tail among the other three solutions but had a bigger IQR.
The scatterplot Figure \ref{fig: pid2dice} shows the data distribution of patient ID versus the average dice score on the BraTS2020 training dataset. The average dice score is obtained by adding dice scores from WT, CT, and ET divided by three. The average dice score is multiplied by $100$ to scale the initial range of $0$ to $1$ to a range of $0$ to $100$. The results are obtained from the online evaluation engine provided by the CBICA Image Processing Portal of the University of Pennsylvania. The scatter plot shows that the dice score drops between patient ID $260$ to $335$. These patient data might be from a different distribution than the other images.

Additionally, data level analysis found that some patients in the mentioned range do not have ET regions. If the model happens to predict even a small ET region, the online evaluation engine added a heavy dice penalty of $0$ and an Hausdorff95 penalty of $373$. A few other patients fall outside the patient ID range of $260$ to $335$ with poor performance. Among the plotted information for the four models, it can be observed that patients trained on weighted focal loss with $C$ = 10 show an improvement compared to the other samples. These are the only information available when analyzing patient ID to average dice score. The general observation and focus will be on data points within the range of patient ID $260$ to about $335$.

\begin{figure}[!hb]
\centering
        \includegraphics[width=0.8\linewidth]{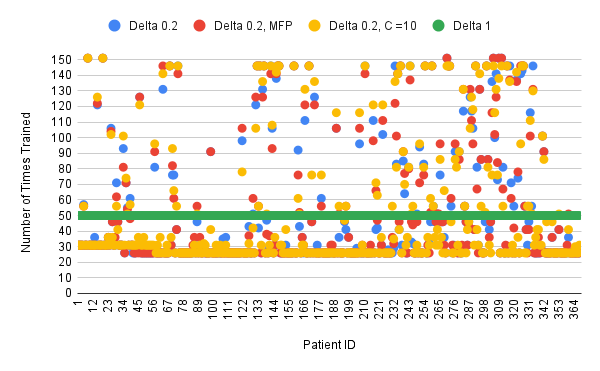}
	\caption{Shows a scatterplot with patient ID vs. the number of times a patient's MR image has been trained.}
	\label{fig: 6pid2trained}
\end{figure}

Figure \ref{fig: 6pid2trained} shows a scatterplot between patient ID and the number of times a patient's MR input image has been trained on the BraTS2020 dataset. Unlike the traditional patient ID vs. average dice score, the new information can provide deeper insights. The former scatter plot can be used to develop future models by studying the data points trained multiple times in the proposed methodology. The regular hybrid MR-U-Net with $\delta$ = 1 will see all the data points equivalent to the total number of epochs. Since the total number of epochs for the hybrid MR-U-Net is $50$, each sample is trained $50$ times denoted by the green scatter points. It is the currently followed methodology and does not provide any additional information. From the other three models trained using dynamic batch training, it can be observed that the model sees a sample at least $25$ times and up to a maximum of $150$ times. The patients that get trained $25$ times are likely easier samples to segment, and those trained to around $150$ times are under-represented or hard samples. It can be observed that a $\delta$ value of 0.2 with the hybrid focal loss with $C$ value of 10 tends to train hard samples more times compared to the other two methods meaning that it gives a higher value to false positives resulting in a higher loss value. Hence, such points get trained multiple times. Compared to the scatter plot in Figure \ref{fig: pid2dice}, the scatter plot in Figure \ref{fig: 6pid2trained} gives more samples to be analyzed as potential outliers.

\subsection{Overall Comparison Study}
Table \ref{tab:mean_validation1} compares various segmentation models on the BraTS2020 validation dataset. The segmentation models include the popular U-Net models and several published works in the literature. The models are evaluated based on two performance metrics: dice Score and Hausdorff95. Models that use post-processing to remove small misclassified tumor regions are not considered for comparison.

\begin{table}[H]
\caption{Shows the comparison study of various models in the literature with the BraTS2020 validation dataset.}
\resizebox{\textwidth}{!}{
\begin{tabular}{cllllll}
\hline
                                         & \multicolumn{3}{c}{Dice Score  $\uparrow$}                                                                                                                                               & \multicolumn{3}{c}{Hausdorff95 $\downarrow$}                                                                                                                 \\ \cline{2-7} 
\multirow{-2}{*}{Model}                  & \multicolumn{1}{c}{ET}                                                          & \multicolumn{1}{c}{WT}                                                         & \multicolumn{1}{c}{TC}                                                          & \multicolumn{1}{c}{ET}                                                          & \multicolumn{1}{c}{WT}                                     & \multicolumn{1}{c}{TC}                                                 \\ \hline
U-Net \cite{ronneberger2015u}             & { 0.688}          &  {0.875} &  0.649          &  {36.15} &  15.15          &  26.90       \\

V-Net \cite{milletari2016v}               &  0.664           &  {0.873} &  0.673          &  51.19          &  15.94          &  24.69                                                                                    \\
3D U-Net \cite{cciccek20163d}               &  \textbf{0.729}           &  0.851 &  \textbf{0.779}          &  \textbf{31.69}          &  12.50          &  18.75                                                                                   \\

Attention U-Net \cite{oktay2018attention} &  0.646                              & 0.863                     & 0.679                     &  56.37                             & 16.81                     &  27.09                \\

U-Net 3 Plus \cite{huang2020unet}         & 0.663                     & 0.868                     &  0.665                              & 48.21                     & 14.50                              & 26.47                \\

MTAU \cite{awasthi2021multi} & 0.570                     & 0.730                     &  0.610                              & 38.87                     & 20.81                              & 24.22                \\

Probabilistic U-Net \cite{savadikar2021brain}  & 0.689                     & 0.819                     &  0.717                              & 36.89                     & 41.52                              & 26.28                \\
DRU-Net \cite{colman2021dr} & 0.670                     & \textbf{0.880}                     &  0.670                              & 47.62                     & 12.11                              & 15.74                \\

MVP U-Net\cite{zhao2021mvp} & 0.670                     & 0.860                     &  0.620                              & 47.33                     & 12.58                              & 50.14                \\

3D U-Net \cite{ballestar2021mri} & 0.720                     & 0.830                     &  0.770                              & 37.42                     & 12.34                              & 13.11                \\

Swin U-Net  \cite{cao2023swin}            & 0.641                    & 0.860                     & { 0.682}                     & { {43.57}}                     & { 11.13}                     & { 16.95}                                                                            \\ \hline

Hybrid MR-U-Net \cite{sahayam2022brain}              &  0.691                              & 0.876                     &  0.644                              &   34.43               &  {14.58}                     &  32.93 \\

Hybrid MR-U-Net Edge                           &  0.681                  &  0.847                            & 0.693                    & 40.70                            & 19.76                             & 20.88                                                                         \\

    Hybrid MR U-Net MFP Focal Loss                               & 0.674                  & 0.870                  & 0.676                  & 38.66                  & \textbf{8.43}                   & 17.43   
    \\
    Hybrid MR U-Net Hybrid Focal Loss               & 0.674                  & 0.874                  & 0.710                  & 41.62                  & 8.81                   & \textbf{12.81}                  \\
    \hline

\end{tabular}
}
\label{tab:mean_validation1}
\end{table}

When comparing the dice coefficient, 3D U-Net models by \cite{cciccek20163d} and \cite{ballestar2021mri} have achieved the best and similar scores for ET and TC regions. The performance is due to the incorporation of 3D MR brain volumes capturing the spatiotemporal information, unlike 2D MR brain regions in which the spatiotemporal connectivity information is usually lost. The problem with 3D deep learning models is the number of parameters required will be significantly larger than 2D deep learning models. Besides the 3D U-Nets, the hybrid U-Net \cite{sahayam2022brain} has obtained the best results for the ET tumor region. For the TC region, the next best performance is by the probabilistic model by \cite{savadikar2021brain}, and the hybrid MR U-Net with hybrid focal loss ($C$ = 10) achieved equivalent results. In the whole tumor region, U-Net, V-Net, DRU-Net, hybrid MR U-Net, and both the hybrid U-Net with the proposed focal achieved comparable results in the dice score.

In the Hausdorff95 metric, the hybrid U-Net with the proposed focal loss achieved significant results compared to other models in the literature. The models performed better than even 3D U-Net. The hybrid MR U-Net obtains the best HD95 result for the TC core with hybrid focal loss ($C$ = 10).  It shows that the training of hard samples with a focus on false positives results in the reduction of small tumor regions without relying on external post-processing. However, the performance on the ET region remains poor, and the 3D U-Net model by \cite{cciccek20163d} achieves the best overall result.

\section{Discussion and Conclusions}
\label{chap: conclusion}

In recent years, there has been an increasing interest in developing accurate brain tumor segmentation models to diagnose and treat brain cancer. This paper proposes a dynamic batch training method to improve the performance of brain tumor segmentation. The proposed method employs two loss functions, hybrid focal loss, and average false positive focal loss, to deal with the problem of class imbalance arising from the absence of tumor regions in certain classes and to capture false positives in the absence of a tumor region.  An ablation study has been performed to identify and study the proposed method using different values of $\delta$ for selecting dynamic batches and $C$ for the hybrid focal loss to obtain a good performance. Various popular encoder-decoder models like U-Net, V-Net, Attention U-Net, U-Net 3 Plus, Swin U-Net, and several models in the literature are compared. The results are compared in Table \ref{tab:mean_validation1}.

In deep learning, identifying hard samples can help improve the model's performance. One approach to identifying hard samples is to keep track of the number of times a patient has been trained. The idea is that samples repeatedly misclassified by the model may be considered hard samples. By studying these hard samples, researchers can gain insight into the model's limitations and develop more efficient deep-learning models in the future. A scatter plot shown in Figure \ref{fig: pid2dice} has been created between patient ID and dice score, and Figure \ref{fig: 6pid2trained} shows a scatter plot between patient ID and the number of patients. It shows that the proposed dynamic batch training can identify under-represented and hard samples in a given dataset. From the results obtained, the following observations and conclusions have been made,

\begin{itemize}
\setstretch{1.5}
\item The dynamic batch training method, in combination with hybrid focal loss and means false positive focal loss, effectively addresses the issue of class imbalance and captures false positives in the absence of a tumor region. It can be observed that a $\delta$ of 0.2 performs well with the hybrid focal loss with $C$ = 10. The experimental results show that the proposed method significantly outperforms the several models in the literature and popular U-Net models in terms of Hausdorff95 distance in the WT and TC metrics on the BraTS2020 dataset.

\item The proposed model's ability to effectively handle false positives and hard samples makes it particularly relevant for clinical settings. Accurate segmentation of brain tumors is crucial for precise diagnosis and treatment planning, and the proposed model shows promise in this regard.

\item Identifying hard samples can also help researchers identify new features and under-represented data that can be used to improve the model's performance and can act as a tool for data collection focusing on patient data that are trained for more time. By focusing on these hard samples, researchers can fine-tune the model to handle complex cases better, ultimately improving overall performance on the task at hand. 

\item The identification and handling of hard samples through dynamic batch training contribute to improved model performance. By focusing on hard samples, researchers can gain valuable insights into the model's limitations and develop more advanced deep-learning models in the future.
\end{itemize}

In conclusion, the proposed model incorporating dynamic batch training, hybrid focal loss, and average false positive focal loss offers significant improvements in brain tumor segmentation. The study demonstrates its superiority over existing models, highlighting its potential for accurate and efficient segmentation of brain tumors. Some of the limitations and future directions are listed below,

\begin{itemize}
\setstretch{1.5}
    \item Even though the Hausdorff95 results have improved for WT and TC regions, there are further chances for improvement for HD95 for the ET region.

    \item Models do not consider the 3D nature of the MR images. Future works could focus on 3D deep learning models that process data with comparatively fewer parameters than 3D U-Net.

    \item Model scalability remains a challenge. However, the proposed model with minor variation leads to performance similar to U-Net and similar architectures with more parameters.

    \item Future work could also propose solutions to improve dice tumor core and enhance tumor metrics in the validation phase.

    \item The batches need to be frozen unlike the traditional batching and each batch should contain similar or temporally related information. The proposed method could give good results to video and 3D image data that satisfies this constraint and can be a future direction.

    \item MR images are 3D in nature. There is a potential to identify novel solutions to implement the proposed method for 2D unrelated images, unlike continuous related 2D brain image slices of a 3D MR image.

    \item Explore the identification of outliers and hard samples by studying the number of times a sample is passed to the model when dynamically trained. The insights can be used to develop novel deep-learning models, improve the explainability of the results, and identify problems, outliers, under-represented samples, and complex latent representations.

    \item Development of a software tool and deployment in a clinical setting needs to be explored in the future. The developed models should be capable of running in low-computing facilities in hospitals and should provide results in a reasonable amount of time.
    
\end{itemize}

\bibliographystyle{elsarticle-num}
\bibliography{reference}

\end{document}